\documentclass[preprint,showpacs,preprintnumbers,amsmath,amssymb]{revtex4}

\usepackage{graphicx}
\usepackage{dcolumn}
\usepackage{bm}

\begin{document}

\def\be{\begin{equation}}
\def\ee{\end{equation}}
\def\bea{\begin{eqnarray}}
\def\eea{\end{eqnarray}}

\preprint{ } \vskip .5in
\title{Classical stability of stringy wormholes in flat and AdS spaces }

\author{Jin Young Kim\footnote{Electronic address:
jykim@kunsan.ac.kr}}
\address{Department of Physics, Kunsan National University,
Kunsan 573-701, Korea}
\author{Yoonbai Kim \footnote{Electronic address:
yoonbai@skku.ac.kr}}
\address{BK21 Physics Research Division and Institute of Basic
Science, Sungkyunkwan University,~Suwon 440-746, Korea\\
School of Physics,~Korea Institute for Advanced Study, 207-43,
Cheongryangri-Dong,~Dongdaemun-Gu, Seoul 130-012, Korea}
\author{James E. Hetrick \footnote{Electronic address:
jhetrick@uop.edu}}
\address{Department of Physics, University of the Pacific,
3601 Pacific Avenue, Stockton, CA 95211}

\date{\today}

\begin{abstract}
We study small fluctuations of the stringy wormhole solutions of
graviton-dilaton-axion system in arbitrary dimensions. We show
under O($d$)-symmetric harmonic perturbation that the Euclidean
wormhole solutions are unstable in flat space irrespective of
dimensions and in anti de Sitter space of $d=3$.
\end{abstract}

\maketitle

\section{Introduction}
Wormholes can play role in many interesting phenomena in quantum
gravity. It is known that there are various kinds of Euclidean
wormhole solutions. In four dimensions the following matter fields
were adopted to support the throat of the wormhole, e.g., axion
field~\cite{gidstr}, scalar fields~\cite{Leecole}, SU(2)
Yang-Mills fields~\cite{hosoya}. Higher dimensional wormhole
solutions were obtained~\cite{myers}, and a higher derivative
correction to the Einstein-Hilbert action was
considered~\cite{fukuta}. Wormhole solutions based on string
theory were studied in Ref.~\cite{gid, rey} where massless
dilaton, in addition to the axion, was naturally included. For
stringy wormholes, the existence of non-singular solutions with
finite action depends on how the dilaton couples to the axion.
Recently Gutperle and Sabra constructed instanton and wormhole
solutions in dilaton-axion-graviton sector of $d$-dimensional
supergravity theory induced from string theory~\cite{gutper}. They
found the critical value for coupling below which non-singular
solutions exist.

Wormhole configurations make non-trivial contributions to the
functional integral in quantum gravity. Wormholes are usually
assumed to make real contribution to amplitudes in quantum
gravity. However, Rubakov and Shvedov~\cite{rubshe} found a
negative mode among fluctuations about the Giddings-Strominger
wormhole solution~\cite{gidstr} in the context of semiclassical
approach, which implies that the wormhole contribution to the
Euclidean functional integral involves imaginary contribution.
This suggests the interpretation of the wormhole as describing the
instability of a large universe against the emission of baby
universes. Similar analysis of the stringy wormholes were carried
out~\cite{klm} under the assumption of a linear relation between
perturbed fields. They confirmed the existence of continuous
spectrum of the negative modes about the nonsingular wormhole
background.

In this paper, we will consider small fluctuations around the
wormhole solutions in $d$-dimensions. To look into the signature
of lowest modes, we use $s$-wave perturbation with O($d$) symmetry
and obtain coupled linear equations. Similar to the stability
analysis of black holes~\cite{chandra}, harmonic form of
perturbation is assumed and it provides the algebraic constraint
for the stability, depending on couplings, the value of
cosmological constant, spacetime dimensions, and integration
constants. It is confirmed that the stringy Euclidean wormhole
solutions include instability for flat space in arbitrary
dimensions irrespective of the value of integration constant and
three-dimensional anti de Sitter space with negative integration
constant where closed form of the wormhole solutions were
obtained.

The organization of the paper is as follows. In Sec. II we briefly
review the wormhole solutions in $d$-dimensions. With keeping in
mind the analysis of small fluctuations, we perform our
calculation with nontrivial lapse function and choose a gauge when
we start to solve the equations. In Sec. III we consider small
fluctuations of O($d$) symmetry and obtained coupled equations of
motion for the perturbed fields. Assuming the harmonic
perturbation, we find the criterion for the stability and apply
this criterion for some special cases. Finally we conclude and
discuss our results in Sec. IV.

\section{Wormhole solutions}

In this section we recapitulate the instanton and wormhole
solutions obtained in Ref.~\cite{gutper}. Bosonic sector of
effective supergravity action of type II string theories involves
a dilaton $\phi$ and an axion $\chi$ in addition to graviton
$g_{\mu\nu}$ while two second-rank antisymmetric tensor fields and
self-dual fourth-rank tensor field are turned off. This
graviton-dilaton-axion system has been taken into account for
various purposes~\cite{gid,rey} and and its generalization to
arbitrary dimensions was performed for obtaining instanton and
wormhole solutions~\cite{gutper}.

Let us begin with the action of the bosonic sector of type II
theory. In string frame it is described by \be S_{\rm S} = \int
d^d x \sqrt{-g_{\rm S}} \left [ e^{-2 \phi} ( R + 4 \nabla_\mu
\phi \nabla^\mu \phi ) - {1 \over {2 (d-1)!} } F_{d-1}^2 - V_{\rm
S} (\phi) \right ], \label{actstfr} \ee where $F_{d-1}$ is an RR
$(d-1)$-form field strength and the subscript ${\rm S}$ denotes
the string frame being used. Note that the dilaton potential is a
product of nonperturbative effects so we consider only flat
potential (or equivalently cosmological constant) in this paper.
Throughout a scale transformation $g_{{\rm S}\mu\nu} = e^{{4 \over
{d-2}} \phi} g_{\mu\nu}$ with $\sqrt{8/(d-2)} \phi \rightarrow
\phi$, the action (\ref{actstfr}) is written in the Einstein frame
as \be\label{act} S = \int d^d x \sqrt{-g} \left [ R - {1 \over 2
} (\nabla \phi)^2 - {1 \over 2} e^{\sqrt{(d-2)/2} \phi} {1 \over {
(d-1)!} } F_{d-1}^2 - V_{\rm E}(\phi) \right ]. \ee Performing a
duality transformation from the $(d-2)$-form antisymmetric tensor
field to the axion field $\chi$ such as $d \chi =
e^{-\sqrt{(d-2)/2} \phi} *F_{d-1}$ in the context of path integral
formalism~\cite{Leecole}, we arrive at the action of our interest
in Minkowski signature \be S_{\rm M} = \int d^d x \sqrt{-g} \left
[ R - {1 \over 2 } (\nabla \phi)^2 - {1 \over 2} e^{ b \phi}
(\nabla \chi)^2 - V(\phi) \right ], \label{acteires} \ee where
$b=\sqrt{(d-2)/d}$ and thus $b= 2$ in ten-dimensional type IIB
theory.

Upon analytic continuation to Euclidean space, the kinetic term of
the axion changes sign in the action (\ref{acteires}) \be S_{\rm
E} = \int d^d x \sqrt{g} \left [ R - {1 \over 2} (\nabla \phi)^2 +
{1 \over 2} e^{b \phi} (\nabla \chi)^2 - V (\phi) \right ].
\label{acteiaxi} \ee From here on let us look for the
O($d$)-symmetric wormhole solution by considering $\chi$ as the
matter field which supports throat of the wormhole. In general the
potential $V (\phi)$ is a function of the dilaton field $\phi$,
which involves nonperturbative contribution of string theory. Here
we take into account only the flat potential independent of
$\phi$, corresponding to a cosmological constant $\Lambda=V$.
Since the dilaton-axion system is invariant under a global U(1)
transformation, we have a conserved current $j_\mu = e^{b \phi}
\nabla_\mu \chi $; \be \nabla_\mu j^\mu = 0. \label{conscur} \ee
Note that this current conservation is paraphrased by the Bianchi
identity of RR$(d-1)$-form field strength in the
duality-transformed theory.

First of all, we introduce the most symmetric O($d$)-invariant
ansatz for the metric \be d s^2 = n^2 (r) dr^2 + a^2 (r) d
\Omega_{d-1}^2 . \label{odmet} \ee Subsequently we demand that
$\phi$ and $\chi$ depend only on the radial coordinate $r$. Then
the O($d$)-symmetric current density $j^0 (r)$ is the only
nonvanishing component and its conservation (\ref{conscur}) lets
$\sqrt{g} j^0$ be a constant. In fact, this constant leads to the
constant global U(1) charge of the wormhole, \be \sqrt{g} j^0 = n
a^{d-1} j^0 = i q , \label{charge} \ee where $i$ results from the
rotation of the action to Euclidean signature which coincides with
the formal replacement $\chi\rightarrow i\chi$.

{}From Eq.~(\ref{charge}), the axion $\chi$ is expressed in terms
of the dilaton $\phi$ and the metric functions $n$ and $a$: \be
\partial_r \chi ={ { i q n e^{- b \phi} } \over a^{d-1} } .
\label{chieq} \ee With the metric ansatz (\ref{odmet}), the
nonvanishing components of Ricci tensor are given as \bea
R_{rr} &=& - (d-1) { {n a'' - n' a'} \over na } , \label{riccirr} \\
R_{ij} &=& - { { n a a'' - n' a a' + (d-2) n {a'}^2 - (d-2) n^3 }
\over n^3} \delta_{ij} , \label{ricciij} \eea where the prime
($'$) denotes the differentiation with respect to $r$.
Substituting Eqs.~(\ref{chieq})--(\ref{ricciij}) into the action
(\ref{acteiaxi}), we obtain the following one-dimensional action
after integrating out the angular variables \bea S_{\rm E} &=&
{\rm Vol}(S_{d-1}) \int d r \Big [ (d-1)(d-2) { { a'^2 a^{d-3} }
\over n }
+ (d-1)(d-2) n a^{d-3} \nonumber \\
&&\hspace{20mm} - { 1 \over 2} { a^{d-1} \over n } {\phi^\prime}^2
- {1 \over 2} { {q^2 e^{-b \phi} n } \over a^{d-1} } - \Lambda n
a^{d-1} \Big ]. \label{actoned} \eea From the action
(\ref{actoned}), we read the equations for $\phi$, $n$ and $a$ \be
\phi: ~~~~~~~~~~~~~
\partial_r \Big ( { a^{d-1} \over n } \partial_r \phi
\Big ) + {bq^2 \over 2} {n \over a^{d-1}} e^{-b \phi} = 0,
~~~~~~~~~~~~~~~~~~~~~~~~~~~~~~~~~~~~ \label{dilem} \ee \bea n:
~~~~~~~~ &-& (d-1)(d-2) { {(\partial_r a)^2 a^{d-3}} \over n^2} +
(d-1)(d-2) a^{d-3} + {1 \over 2} { a^{d-1} \over n^2} (\partial_r
\phi)^2
\nonumber \\
&-& {1 \over 2} { { q^2 e^{- b \phi} } \over a^{d-1} } - \Lambda
a^{d-1} = 0, \label{nem} \eea \bea a:~~~~~~~~~~~~ && \partial_r
\left[ 2(d-2) { {(\partial_r a) a^{d-3} } \over n } \right] -
(d-2)(d-3) \left[ { { (\partial_r a)^2 a^{d-4} } \over n }
+ n a^{d-4} \right] \nonumber \\
&&+ {1 \over 2} { a^{d-2} \over n} (\partial_r \phi)^2 - {1 \over
2} { { q^2 n e^{- b \phi} } \over a^{d} } - \Lambda n a^{d-2} = 0.
\label{aem} \eea We set the lapse function to be $n=1$ as a
gauge-fixing condition, which is one of admissible gauges for our
purpose~\cite{TS}. Then the dilaton equation of motion
(\ref{dilem}) can be integrated by multiplying $a^{d-1} \partial_r
\phi$ \be (\partial_r \phi)^2 - {q^2 \over a^{2d-2} } e^{-b \phi}
- { c \over a^{2d -2} } = 0 , \label{dileqc} \ee where $c$ is an
integration constant. Using this integral, the gravitational
equations (\ref{nem}) and (\ref{aem}) give \be 1 - (\partial_r
a)^2 + {c \over {2(d-1)(d-2) a^{2d-4} } } - {\Lambda \over
{(d-1)(d-2) } }a^2 = 0. \label{clameq} \ee The equation
(\ref{clameq}) can be solved to find $a(r)$ and then one can find
$\phi (r)$ from Eq.~(\ref{dileqc}) and $\chi(r)$ from
Eq.~(\ref{chieq}). The solution will depend on the choice of
$\Lambda$ and $c$. The cases $\Lambda = 0$, $\Lambda < 0$ and
$\Lambda > 0$ correspond to asymptotically flat, anti-de Sitter,
and de Sitter space, respectively. The solution for $c \ne 0$
corresponds to a wormhole while the one for $c=0$ corresponds to
an instanton. Since we are interested in the stability of the
wormhole solution, we consider the $c \ne 0$ case. Obviously the
same equations in Eqs.~(\ref{dileqc}) and (\ref{clameq}) can be
obtained under the gauge-fixed metric~\cite{gutper}: \be d s^2 =
dr^2 + a^2 (r) d \Omega_{d-1}^2 , \ee however until now we keep
the gauge degree $n(r)$ for the analysis of small fluctuations.

The wormhole solution is obtained when the metric function $a(r)$
has a minimum value $a_0$ at the neck of the wormhole with
$\partial_r a (r) = 0$. From Eq.~(\ref{clameq}), we have \be
\pm\int^{a}_{a_{0}} \frac{da}{\sqrt { 1 + {c \over {2(d-1)(d-2)
a^{2d-4} } } - {\Lambda \over {(d-1)(d-2) } }a^2 }} =r,
\label{wormha} \ee where we choose the position of the neck as the
origin, $a(0)=a_{0}$, without loss of generality. This Euclidean
configuration can be continued to Minkowski space. The dilaton can
be solved from Eq.~(\ref{dileqc}) \be \int { {d \phi} \over \sqrt{
q^2 e^{-b \phi} + c } } = \pm \int {dr \over a^{d-1} } .
\label{dilsolgen} \ee Finally, from Eq.~(\ref{chieq}), the axion
configuration is given by \be i(\chi(r)-\chi_{0})=-q\int dr
\frac{e^{-b\phi}}{a^{d-1}}. \ee

The solutions for possible values of $\Lambda$ and $c$ for
arbitrary dilaton coupling $b$ were found recently in
Ref.~\cite{gutper}. The characters of background solution
necessary for our stability analysis can be summarized as follows.
For $\Lambda = 0$, one has a minimal size sphere when $c < 0$. The
neck of the wormhole and the dilaton profile are given by \bea a_0
&=& \left[ { {2 (d-1) (d-2)} \over {|c|} }
\right]^{- {1 \over {2d -4} } } , \label{azero} \\
e^{ {b \over 2} \phi_0 } &=& { {|q|} \over {|c|^{1 \over 2} } }
\sin \left[ \sin^{-1} \Big( \sqrt{ {|c|} \over q^2} e^{ {b \over
2} \phi_\infty } \Big) + {\pi \over 2 } |b| \sqrt{ {d-1} \over
{2(d-2) }} \, \right] , \label{phizero} \eea where
$\phi_{\infty}=\phi(r=\infty)$. For $\Lambda < 0$, wormhole
solution can exist for $c < 0$. The solution can be expressed in
terms of elementary functions for the case $d = 3$ \bea a_0^2 &=&
{ {\sqrt{ 1 + { {|\Lambda c| } \over 2} } - 1}
\over {|\Lambda|} } , \label{alaml0}\\
e^{ {b \over 2} \phi_0 } &=& { {|q|} \over {|c|^{1 \over 2} } }
\sin \left[ \sin^{-1} \Big ( \sqrt{ {|c|} \over q^2} e^{ {b \over
2} \phi_\infty } \Big ) \mp { {|b|} \over 2} \left( {\pi \over 2 }
+ \sin^{-1} \Big ( { 1 \over \sqrt {1 + { {|c|\Lambda} \over 2} }
} \Big ) \right) \right] . \label{philaml0} \eea The solutions for
$d = 4, 5$ can be solved in terms of elliptic integrals. When
$c>5$, no solution in closed form is reported by analytic method,
yet. For $\Lambda > 0$, there is no non-singular solution. So we
will not consider this case anymore.

\section{Small fluctuations and stability analysis}

We consider small fluctuations with O($d$) symmetry about the
obtained nonsingular wormhole solution. Since the important issue
in Euclidean wormhole physics is to decide whether the Euclidean
wormhole configurations can have purely real contributions to the
functional integral or include imaginary contributions, simple
study of $s$-wave perturbation can precede complicated systematic
analysis of small fluctuations with angle dependence in this
semiclassical approach~\cite{rubshe}. To be specific, let us
consider \bea
n(r) &=& 1 + {\tilde n} (r) , \\
a(r) &=& a_0 + {\tilde a} (r) , \\
\phi (r) &=& \phi_0 + {\tilde \phi} (r), \eea where $a_0$ and
$\phi_0$ are given in the wormhole solutions obtained from
Eqs.~(\ref{dileqc}) and (\ref{clameq}). Substitute these into
Eq.~(\ref{actoned}) and take only the bilinear terms in $({\tilde
n},{\tilde a},{\tilde \phi})$ of the action. From these terms one
can derive the linearized equations for the small fluctuations. In
order to keep the consistency with the gauge-fixing condition
$n=1$ we forced, here we also take ${\tilde n} = 0$. The bilinear
action is calculated as \be S_{\rm bil} = {\rm Vol}(S_{d-1}) \int
d r a_0^{d-1} \Big[ A_0 {\tilde a}^{\prime 2} + B_0 {\tilde
\phi}^{\prime 2} + C_0 {\tilde a} {\tilde \phi}^\prime + D_0
{\tilde a} {\tilde \phi} + E_0 {\tilde a}^2 + F_0 {\tilde \phi}^2
\Big ] , \ee with boundary terms which are not relevant for our
analysis. Here $A_0 , \cdots , F_0$ are given as \bea
A_0 &=& { {(d-1)(d-2)} \over a_0^2 }, \nonumber \\
B_0 &=& - {1 \over 2} , \nonumber \\
C_0 &=& - { {(d-1)} \over a_0 } \phi_0^\prime , \nonumber \\
D_0 &=& - {1 \over 2} { {(d-1)} \over a_0 } b Q^2 , \nonumber \\
E_0 &=& { {(d-1)(d-2)(d-3)(d-4)} \over 2 } {1 \over a_0^4 } - { {d
(d-1)} \over {4 a_0^2} } Q^2 - { {(d-1)(d-2)} \over 2 } \Big ( { {
{\phi_0}^\prime}^2 \over {2 a_0^2} }
+ { { \Lambda} \over a_0^2 } \Big ) , \nonumber \\
F_0 &=& - { b^2 \over 4 } Q^2 , \eea where \be Q^2 = { {q^2 e^{- b
\phi_0}} \over a_0^{2(d-1)} } . \label{defq} \ee

The equations of motion for ${\tilde a}$ and $ {\tilde \phi}$ are
summarized by \bea && - {\tilde a}^{\prime\prime} + { C_0 \over
2A_0} {\tilde \phi}^\prime + { E_0 \over A_0} {\tilde a} + { D_0
\over 2A_0} {\tilde \phi}= 0 ,
\label{fla}\\
&& - {\tilde \phi}^{\prime\prime} + C_0 {\tilde a}^\prime - D_0
{\tilde a} - 2 F_0 {\tilde \phi} = 0 . \label{flp} \eea Since they
are coupled linear differential equations of second-order, we
employ the assumption of harmonic form which looks appropriate for
testing existence of negative modes \be {\tilde a} = {\tilde a}_0
e^{i \omega r} , ~~~~~~ {\tilde \phi} = {\tilde \phi}_0 e^{i
\omega r} . \nonumber \ee Then the normal modes of the coupled
equations are determined by a $2\times 2$ matrix equation \be
\begin{pmatrix} \omega^2 + {E_0 \over A_0}&
{ 1 \over {2 A_0} } ( D_0 + i \omega C_0 ) \cr - ( D_0 - i \omega
C_0 ) & \omega^2 - 2 F_0 \cr
\end{pmatrix}
\begin{pmatrix}
{\tilde a}_0 \cr {\tilde \phi}_0 \cr
\end{pmatrix}
~=~0. \ee From the determinant of the matrix equation we obtain
\be \omega^4 + \alpha \omega^2 + \beta = 0 , \label{omegaeq} \ee
where \be \alpha = { E_0 \over A_0} - 2 F_0 + {C_0^2 \over {2 A_0}
} , ~~~~ \beta = { D_0^2 \over {2 A_0} } - { {2E_0 F_0} \over {2
A_0} } . \ee The explicit forms of $\alpha$ and $\beta$ are
calculated as, upon eliminating $\phi_0^\prime$ using
Eq.~(\ref{dileqc}), \bea \alpha &=& { {(d-3)(d-4)} \over {2 a_0^2
} } + {1 \over 4} { d \over {d-2} } { c \over a_0^{2(d-1)} }
+ {b^2 \over 2} Q^2 - { \Lambda \over 2 } , \label{alphaexp} \\
\beta &=& {b^2 \over 2 } Q^2 \Big ( { {(d-3)(d-4)} \over {2 a_0^2
} } -{1 \over 4} { c \over a_0^{2(d-1)} } -{1 \over 4} { {d-1}
\over {d-2} } Q^2 - { \Lambda \over 2 } \Big ) . \label{betaexp}
\eea The condition for the wormhole solution not to have a
negative mode over the perturbation is that Eq.~(\ref{omegaeq})
has all real roots. The existence of imaginary part in $\omega$
means that $e^{i \omega r}$ can grow exponentially. This tells
instability of the solution. The condition for $\omega$ to have
all real roots is equivalent to $\omega^2$ to have all
non-negative real roots. So the criterion for the stability can be
written as \be \alpha < 0 ,~~~~ \beta > 0, ~~~~ \alpha^2 - 4 \beta
> 0 . \label{stacri} \ee

\subsection{Flat space}
In this case, it has been pointed out that the wormhole solution
exists for $c < 0$~\cite{gutper}. Taking $\Lambda = 0 $ in
Eqs.~(\ref{alphaexp}) and (\ref{betaexp}), the conditions in
Eq.~(\ref{stacri}) become \bea { {- 3(d-3)} \over {a_0^2 } }
+ {{b^2} \over 2} Q^2 &<& 0, \label{alphaflat} \\
{ {b^2 Q^2 } \over 2 } \Big ( { {d^2 - 5d +7} \over { a_0^2 } }
-{1 \over 4} { {d-1} \over {d-2} } Q^2
\Big ) &>& 0 , \label{betaflat} \\
{b^2 \over 2} \Big ( {b^2 \over 2} + { {d-1} \over {d-2} } \Big )
Q^4 - b^2 (2 d^2 -7 d +8) {Q^2 \over a_0^2} + { {9(d-2)^2} \over
a_0^4 } &>& 0 . \eea We restrict ourselves to the type II case
where $b$ is given by $b= \sqrt{(d-2)/2}$ for simplicity. By
solving the above inequalities in terms of $Q$, we obtain a
constraint on the parameters for stability condition \bea
d=3: ~~~~~~ && Q^2 < {8 \over a_0^2} , \\
d=4: ~~~~~~ && Q^2 < {2 \over a_0^2} , \\
d \ge 5: ~~~~~~&& Q^2 < { {4(d-2)} \over {d^2 a_0^2} } ( 2d^2 -7d
+8 - \sqrt{(2d^2 -7d +8 )^2 - 9 d^2} ). \eea Using
Eqs.~(\ref{azero}), (\ref{phizero}) and (\ref{defq}), the
stability condition is expressed in terms of U(1) charge $q$ and
the parameter of wormhole solution $c$ \bea d=3: ~~&&\sin \Big (
\sin^{-1} \Big ( \sqrt{ {|c|} \over q^2} e^{ {b \over 2}
\phi_\infty } \Big ) + { {\sqrt{2} \pi} \over 4 } \Big ) >
\sqrt{2} ,
\label{flatde3} \\
d=4: ~~&&\sin \Big ( \sin^{-1} \Big ( \sqrt{ {|c|} \over q^2} e^{
{b \over 2} \phi_\infty } \Big ) + { {\sqrt{3} \pi} \over 4 } \Big
) > \sqrt{3 \over 2},
\label{flatde4} \\
d \ge 5: ~~ && \sin \Big ( \sin^{-1} \Big ( \sqrt{ {|c|} \over
q^2} e^{ {b \over 2} \phi_\infty }
\Big ) + { {\sqrt{d-1} \pi} \over 4 } \Big ) \nonumber \\
&& > \Big [ {{d-1} \over 18} \Big ( 2d^2 -7d +8 + \sqrt{( 2d^2 -7d
+8)^2 - 9 d^2} \Big ) \Big ]^{1 \over 2} . \label{flatdge5} \eea
The right-hand side of Eq.~(\ref{flatdge5}) is larger than one for
$d \ge 5$. Therefore Eqs.~(\ref{flatde3}), (\ref{flatde4}) and
(\ref{flatdge5}) cannot hold for any choice of $c$ and $q$. Thus
we conclude that the wormhole solution with $\Lambda =0$ shows
unstable behavior under the small fluctuations of O($d$) symmetry.

\subsection{Anti de Sitter space}

For $\Lambda <0$, wormhole solution can exist for $c < 0$. We
consider the $d = 3$ case where the wormhole solution can be
expressed in terms of elementary functions. Here we also consider
the type II case where $b$ is given by $b = \sqrt{(d-2)/2} $.
Taking $d=3$ and $b=\sqrt{1 /2}$ in Eqs.~(\ref{alphaexp}) and
(\ref{betaexp}), we have \bea \alpha &=& {3 \over 4} { c \over
a_0^4 }
+ {Q^2 \over 4} - { \Lambda \over 2 } , \label{alphaads} \\
\beta &=& {1 \over 4 } Q^2 \Big ( -{1 \over 4} { c \over a_0^4 }
-{1 \over 2} Q^2 - { \Lambda \over 2 } \Big ) . \label{betaads}
\eea From Eq.~(\ref{clameq}), we have for $d=3$ \be 1 + {c \over
{4 a_0^2 } } - {\Lambda \over 2 }a_0^2 = 0. \label{aads3} \ee
Using Eq.~(\ref{aads3}), $\alpha$ and $\beta$ can be simplified
further \bea \alpha &=& -{ 3 \over a_0^2 }
+ {Q^2 \over 4} + \Lambda , \label{alads3} \\
\beta &=& {1 \over 4 } Q^2 \Big ( {1 \over a_0^2} -{1 \over 2} Q^2
- \Lambda \Big ) . \label{beads3} \eea With the above choice of
$\alpha$ and $\beta$, we have \be \alpha^2 - 4 \beta = \Big ( {3
\over 4} Q^2 - {3 \over a_0^2} - \Lambda \Big )^2 + {2 \over
a_0^2} > 0 . \ee The condition $\alpha^2 -4 \beta > 0$ is
satisfied automatically. Thus the constraint for the stability is
obtained, from $\alpha <0$ and $\beta >0$, as \be Q^2 < 2 \Big (
{1 \over a_0^2} + | \Lambda | \Big ) . \label{consads3} \ee Using
Eqs.~(\ref{alaml0}), (\ref{philaml0}) and (\ref{defq}), this
condition can be written as \be\label{sine} \sin \left[ \sin^{-1}
\Big ( \sqrt{ {|c|} \over q^2} e^{ {b \over 2} \phi_\infty } \Big
) \mp {1 \over {2 \sqrt{2} } } \left( {\pi \over 2 } + \sin^{-1}
\Big ( { 1 \over \sqrt {1 + { {|c|\Lambda} \over 2} } } \Big )
\right) \right]
> \Big ( 1 + { 1 \over \sqrt{1 + { {|c \Lambda|} \over 2} } }
\Big )^{1 \over 2} . \ee The right-hand side of Eq.~(\ref{sine})
is always larger than one, so one can conclude that the wormhole
solution is also unstable for three-dimensional anti de Sitter
space.

\section{Discussion}

In this paper we have studied the classical stability of stringy
wormhole solutions. For the small fluctuations with O($d$)
symmetry, the analysis to linear order results in two coupled
differential equations of second order. Under the assumption of
harmonic form perturbation, we obtained a condition for the
criterion of stability of the Euclidean wormhole solutions, which
is a function of the dimensionality of the spacetime, the dilaton
coupling $b$, the cosmological constant $\Lambda$ as well as the
integration constant $c$ distinguishing instanton and wormhole. As
a concrete test we applied our criterion to type II case where the
coupling is given by $b = \sqrt{(d-2)/2}$ and showed that every
wormhole solution expressed in terms of elementary functions is
unstable. It would be also interesting if one could find any
example in string theory where $b$ is in the range where the
wormhole solution is stable under the classical perturbation.

\begin{acknowledgments}
We would like to thank S.P. Kim for useful discussion. J.Y.K.
would like to thank the Department of Physics at University of the
Pacific for hospitality during his visit. This work was supported
by the Korea Research Foundation Grant (KRF-2001-015-DP0082).
\end{acknowledgments}

\end{document}